\documentclass[showpacs,aps,prb,twocolumn,floatfix]{revtex4}
\usepackage{times}
\usepackage{graphicx}
\usepackage{bm}
\usepackage{amsmath}

\newcommand{\s}{\sum\limits}

\newcommand{\be}{\begin{equation}}
\newcommand{\e}{\end{equation}}
\newcommand{\beml}{\begin{subequations}}
\newcommand{\eml}{\end{subequations}}
\newcommand{\beq}{\begin{eqnarray}}
\newcommand{\eq}{\end{eqnarray}}
\newcommand{\ba}{\begin{array}}
\newcommand{\ea}{\end{array}}
\newcommand{\lt}{\left}
\newcommand{\rt}{\right}
\newcommand{\n}{\nonumber}
\newcommand{\la}{\langle}
\newcommand{\ra}{\rangle}

\newcommand{\im}{\,{\rm Im}\,}
\newcommand{\re}{\,{\rm Re}\,}
\newcommand{\ep}{\varepsilon}

\DeclareMathOperator{\var}{var}

\begin{document}
 
\date{February 2003}

\title{Anomalies and non-log-normal tails
in one-dimensional localization with power-law disorder}

\author{M. Titov}
\author{H. Schomerus}

\affiliation{Max-Planck-Institut f\"ur Physik komplexer Systeme,
N\"othnitzer Str. 38, 01187 Dresden, Germany
}

\begin{abstract}
Within a general framework, we discuss the wave function statistics
in the Lloyd model of Anderson localization on a
one-dimensional lattice
with a Cauchy distribution for the random on-site potential.
 We demonstrate that already 
in leading order in the disorder strength,
there exists a 
hierarchy of anomalies in the probability distributions
of the wave function, the conductance, and the local density of states,
for every energy  
which corresponds to a 
rational ratio of wave length to the lattice constant. 
We also show that these distribution functions 
do have power-law rather then log-normal tails 
and do not display universal single-parameter scaling. 
These peculiarities persist in any model with
power-law tails of the disorder distribution function.
\end{abstract}
\pacs{
72.15.Rn,  
05.45.Ac   
}

\maketitle

Wave function  localization in one spatial dimension 
has attracted enormous attention 
since the pioneering work by Anderson.\cite{Anderson} 
Much of our present understanding of this phenomenon
is based on the original Anderson model taken 
on a one-dimensional lattice
\be
\label{AM}
-t(\Psi_{n+1}+\Psi_{n-1})+ V_n \Psi_n=E \Psi_n,
\e
with a white-noise disorder 
$\la V_n V_m \ra \propto \delta_{nm}$,  $\la V_n\ra=0$,
and fixed hopping element $t$. The potential $V_n$ 
at each site
takes real values according to a probability density $P_V(V)$. 
Most of the theoretical investigations of Anderson localization assume
a finite variance $\var V\equiv 2 D <\infty$, and then consider
the weak-disorder limit $D \ll t^2$.
This condition is a prerequisite for single-parameter 
scaling, \cite{ATAF} in which the product
$L \xi^{-1}$
of system length $L$ and inverse localization length
$\xi^{-1}=-\lim_{n\to\infty}\frac{1}{n}\ln |\Psi_n|$
is the only free parameter in  the universal distribution function of
the Lyapunov exponents
$\alpha=-\frac{1}{n}\ln |\Psi_n|$ for finite $n$.
This carries over to universal distribution functions of
the dimensionless conductance $g$ and the local density of states $\nu$.
For $L/\xi\gg 1$ these distribution functions follow log-normal laws ---
the paradigm for large fluctuations.
On the other hand, in many realistic applications
the distribution function $P_V(V)$ displays power-law tails, with $D=\infty$,
such that in a sense disorder never really is weak.
The most prominent example is the localization of wave functions in the
momentum space of the kicked rotator. \cite{FGP} This dynamical problem
has been mapped onto the Anderson model in the seminal works
\onlinecite{FGP,Shepelyanskii}, 
with an effectively random Cauchy-distributed potential
\begin{equation}
\label{cauchy}
P_{V,~\rm Cauchy}(V)=\frac{1}{\pi}\im \frac{1}{V-i\delta},\quad \delta>0
.
\end{equation}
Equation (\ref{AM}) with the disorder distribution function given by
Eq.\ (\ref{cauchy}) has been proposed for the first time by
Lloyd. \cite{Lloyd} In this model the localization length
can be computed analytically for arbitrary $\delta$,
\cite{Thouless,Ishii}
and the variance of
the Lyapunov exponents $\alpha$
has been analyzed very recently in Ref.\ \onlinecite{DLA}.

A beautiful experimental realization of the kicked rotator 
is the dynamics of atoms
driven by a regular train of laser pulses. \cite{Raizen,Raizen2}
In these experiments, the probability distribution function in momentum space
is seen to relax from an initial Gaussian into an exponential profile,
demonstrating the absence of diffusion in momentum direction.
However, since
the wave function statistics in localization is 
a prototypical example of large fluctuations,
the localization length of the eigenstates can be 
quantitatively inferred from moments
of the wave function (like the measured mean probability)
only if the full shape of the
distribution function is known, including its tails. 
The impact of the tails of the probability density
$P_V(V)$ on the statistics of the wave functions 
has been already mentioned by Halperin,\cite{Halperin}
but has not been analyzed, let alone sufficiently appreciated,
in the literature. 

In this paper we address  the fluctuations 
of the wave function $\Psi$ and related quantities 
for distributions $P_V(V)$ with power-law tails,
by going beyond the mean $\xi^{-1}$ of $\alpha$ and its variance,
studied so far. \cite{Thouless,Ishii,DLA}
We first set out a general framework for arbitrary $P_V(V)$,
which then is applied to the Lloyd model
with $P_V(V)$ given by Eq.\ (\ref{cauchy}).
The fluctuations turn out to be highly non-universal,
with an anomalous energy dependence 
[reflecting also the spatial discreetness of the Anderson model (\ref{AM})]
and non-log-normal tails that strongly affect the behavior of the moments
of $\Psi$, $g$, and $\nu$
even in the weak-disorder limit $\delta\ll t$, and even for low
fractional orders of the moments.
These characteristics of  the wave function statistics are
in striking contrast to the universality for models with $D\ll t^2$
and single-parameter scaling. 

The central quantity of interest in our calculation is the 
generating function $\mu(\lambda)$ of the cumulants of $\ln\Psi$.
As pointed out by Borland \cite{Borland} and Thouless,
\cite{Thouless1979} instead of solving 
Eq.\ (\ref{AM}) as a boundary-value problem
it suffices to investigate
the specific solution $\Phi_n$ of the initial-value problem 
$\Phi_0=a$, $\Phi_1=b$.
At large distances $n\gg 1$ this solution exponentially increases as
$\Phi_n \sim \exp(\alpha n)$ for almost all values of $a$ and $b$,
which statistically is equivalent to the inverse wave-function
decay $\Psi_n \sim \Phi_n^{-1}$ in the original problem.
The cumulant-generating function  
\be
\label{definition_of_mu}
\mu(\lambda)=\lim_{n\to\infty} \frac{1}{n}\ln \lt\la 
\lt|\Phi_n\rt|^\lambda \rt\ra=\s_{k=1}^\infty
\frac{c_k}{k!}\lambda^k
\e
accounts for the details of convergence of 
the Lyapunov exponent $\alpha$ to its mean value 
$c_1=\xi^{-1}$. The coefficients $c_k$ with $k\geq 2$ 
are numerical constants which characterize the 
deviations of $\alpha$ from $c_1$.
As follows from Eq.\ (\ref{definition_of_mu}) 
the cumulants of $\alpha$ vanish according to the law 
dictated by the generalized central-limit theorem
\be
\label{clt}
\la\la \alpha^k \ra\ra \sim c_k n^{1-k} \quad \text{for } n\gg\xi. 
\e
The coefficients $c_k$ do not depend  on the initial conditions for $\Phi_n$
and capture the universal information about the fluctuations
of many essential quantities in the localized regime, such as
the dimensionless conductance $g$ of a finite sample of the length $L$
and the local density of states\cite{STBB} $\nu$ in a semi-infinite
wire at a distance $L$ to an open boundary.
Both quantities do strongly fluctuate even in the localized regime 
$L \gg \xi$, unlike the Lyapunov exponent. 
However, the fluctuations of their logarithms can
be expressed through the same coefficients $c_k$ by
\be
\lim_{L\to\infty}\frac{1}{L}\la\la \lt(-\ln g \rt)^k \ra\ra =
\lim_{L\to\infty}\frac{1}{L}\la\la \lt(-\ln\nu\rt)^k \ra\ra = 2^k c_k.
\e
Each coefficient may be used to define a length scale $\xi_k=c_k^{-1}$,
and the question for single-parameter scaling can be posed as 
whether these  length scales are independent quantities or not. \cite{ST2003b}

We now set out a general approach to calculate the
generating function $\mu(\lambda)$ and the coefficients $c_k$
for arbitrary form of $P_V(V)$.
We built up on the formalism previously used to calculate the inverse
localization length $\xi^{-1}=c_1$. \cite{LGP,Derrida,Bovier}
The Anderson model (\ref{AM}) can be written in terms of 
new variables
$z_n=\Psi_{n+1}/\Psi_{n}$, $r_n=\ln|\Psi_n|$
in the simple way
\be
\label{map}
z_n=v_n-1/z_{n-1},\qquad r_n=r_{n-1}+\ln|z_{n-1}|,
\e
where $v_n\equiv (V_n-E)/t$. We seek the specific
solution of the initial-value problem $z_0=b/a$, $r_0=\ln|a|$.
Iterating the map (\ref{map}) 
we observe that $z_n$ and $r_n$ take real values $z$ and $r$
with a probability density $P_n(z,r)$, which obeys 
\beq
P_n(z,r)&=&\int_{-\infty}^{\infty}F(v)\,dv
\int\!\!\!\!\int_{-\infty}^{\infty}
P_{n-1}(z',r')\,dz'dr'\n \\
\label{FPK}
&\times& \delta(z-v+1/z')\,\delta(r-r'-\ln{|z'|}),
\eq
with $F(v)=t P_V(vt+E)$ the probability density of $v$.

It is convenient to introduce the function 
\be
\label{definition_of_h}
h_n(z,\lambda)=|z|^\lambda\int_{-\infty}^\infty
e^{r\lambda}P_n(z,r)\,dr
\e
and rewrite Eq.\ (\ref{FPK}) as
\be
\label{hEq}
h_n(z,\lambda)=|z|^\lambda\int_{-\infty}^\infty F(z+1/z')\,
h_{n-1}(z',\lambda)\,dz'.
\e
According to Eqs.\ (\ref{definition_of_mu}) and 
(\ref{definition_of_h}) we have 
\be
\mu(\lambda)=\lim\limits_{n\to\infty}\frac{1}{n}
\ln\lt[\int_{-\infty}^\infty dz\,h_{n-1}(z,\lambda)\rt]. 
\e
If the function $\mu(\lambda)$ exists, the solution
to Eq.\ (\ref{hEq}) at large $n$  must fulfill the relation
$h_{n}(z,\lambda)=e^{\mu(\lambda)}h_{n-1}(z,\lambda)$.
Thus Eq.\ (\ref{hEq}) is transformed into 
the functional eigenvalue problem
\beml
\label{basic}
\beq
&&e^{\mu(\lambda)-\lambda\ln{|z|}}h(z,\lambda)={\cal F}[h](z,\lambda),\\
&&{\cal F}[h](z,\lambda)\equiv \int_{-\infty}^\infty 
F(z+1/z')\,h(z',\lambda)\,dz'.
\eq
\eml
This is the central general equation of this paper.
In any practical case it
has to be solved perturbatively in $\lambda$. 
We expand the function $h(z,\lambda)$ in a series
\be
h(z,\lambda)=\s_{k=0}^\infty \frac{\lambda^k}{k!}h_k(z)
\e
and introduce the notation
$\tilde{c}_1(z)\equiv c_1-\ln|z|$.
Equation (\ref{basic}) is transformed into the following 
set of equations
\beml
\label{TheSet}
\beq
\label{a}&&{\cal F}[h_0]-h_0=0,\\
\label{b}&&{\cal F}[h_1]-h_1=\tilde{c}_1h_0,\\
\label{c}&&{\cal F}[h_2]-h_2=(\tilde{c}_1^2+c_2)h_0+2\tilde{c}_1h_1,\\
&&{\cal F}[h_3]-h_3=(\tilde{c}_1^3+3\tilde{c}_1c_2+c_3)h_0\n\\
\label{d}&&\qquad\qquad\qquad+3(\tilde{c}_1^2+c_2)h_1+3\tilde{c}_1h_2, 
\mbox{  etc.}
\eq
\eml
Equation (\ref{a}) delivers the stationary distribution
function of $z$ and has been used before \cite{LGP,Derrida,Bovier} 
to calculate 
the localization length $\xi=c_1^{-1}$ from
\be
\label{C1}
c_1=\int_{-\infty}^\infty h_0(z)\,\ln|z|\,dz.
\e
So far we have shown that Eq.\ (\ref{a}) can be considered
as just the first member of a hierarchy of equations that determine the complete
wave function statistics for finite $n\gg \xi$.
 
The integrals $\int_{-\infty}^{\infty}dz$ of the left-hand sides of 
Eqs.\ (\ref{TheSet}) equal zero. 
Equation (\ref{C1}) indeed can be derived by integrating 
both sides of Eq.\ (\ref{b}) along the real axis.
Once the distribution function $h_0(z)$ and the mean Lyapunov exponent $c_1$
are known (as is analytically the case in the Lloyd model),
one can construct the solution to Eq.\ (\ref{b})
iteratively by\cite{footnote}
\beml
\label{kernel0}
\beq
&&h_1(z)={\cal K}[\tilde{c}_1h_0](z),\\
&&{\cal K}[f](z)\equiv -f(z)-\int_{-\infty}^\infty 
K(z,z')\,f(z')\,dz'.
\eq
\eml
The second coefficient $c_2$ in the cumulant expansion
is readily found by integrating Eq.\ (\ref{c}) along the real axis,
\be
\label{C2}
c_2\!=\!\int_{-\infty}^\infty\!\!\! [h_0(z)\,(\ln|z|-c_1)
+2h_1(z)](\ln|z|-c_1)dz.
\e
This procedure can be repeated recursively
to calculate all other coefficients $c_k$. 

\begin{figure}[t]
\includegraphics[width=\columnwidth]{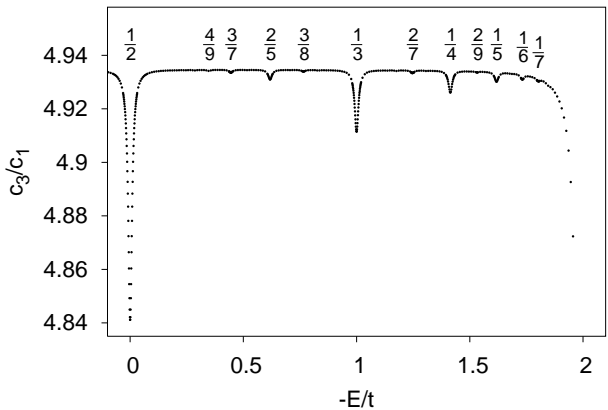}
\caption{The ratio $c_3/c_1$ is plotted according to the analytical
result (\ref{c3}) for the Lloyd model at the disorder strength 
$\delta=0.01 t$.
The ratio is never small inside the band and reveals anomalies
at energies $E=-2 t \cos(\pi p/q)$ with $p$ and $q$ integers. 
The corresponding 
rational number $p/q$ is indicated in the figure.
The size of the anomaly only depends on the value of the denominator $q$.
}
\label{fig:ratio}
\end{figure}

For the  Lloyd model,
the scheme developed above allows us to 
obtain the coefficients $c_k$ analytically. 
The stationary distribution function of the variable $z$
is readily found by the exact solution of the integral equation 
(\ref{a}), \cite{Lloyd,LGP}
\be
h_0(z)=\frac{1}{\pi}\im \frac{1}{z-s},\quad 
s\!+\!\frac{1}{s}=\frac{E\!+\!i\delta}{t},\quad \im s>0.
\e
The integral (\ref{C1}) yields
the well-known result
\be
\label{C1exact}
c_1=\ln|s|.
\e
The kernel function $K(z,z')$ can be 
obtained by iterative application
of the operator ${\cal F}[h]$,
\beml
\label{kernel}
\beq
\label{a1}
&&K(z,z')=\frac{1}{\pi}\im\!\s_{n=1}^\infty\!
\lt(\frac{1}{z\!-\!p_n(z')}-\frac{1}{z\!-\!r_n}
\rt),\\
\label{b1}
&&
p_n(z)=\frac{(s^n-s^{-n})-z(s^{n+1}-s^{-(n+1)})}
{(s^{n-1}-s^{-(n-1)})-z(s^{n}-s^{-n})},\\
\label{c1}
&& r_n=p_n(\pm\infty) 
,
\eq
\eml
where the second term in the parenthesis 
on the right hand side of Eq.\ (\ref{a1})
is added to provide a better convergence 
of the intermediate expressions.\cite{footnote} 

Applying the result (\ref{kernel}) to Eqs.\   (\ref{kernel0}) and (\ref{C2}) 
and performing the summation one recovers the result of Ref.\ \onlinecite{DLA},
\beq
&&c_2=\re\lt[\rm{Li}_2(s^{-2})-\rm{Li}_2(|s|^{-2})\rt]
+\arg(s)(\pi-\arg(s))\n\\
&&\qquad +\ln|s|^2
\lt[ \ln(|s|^2-1)\,-\ln|s^2-1|\rt],
\label{c2}
\eq
where ${\rm Li}_n(z)=\sum_{k=1}^\infty z^k/k^n$ is the 
polylogarithmic function.
As has been shown in Ref.\ \onlinecite{DLA},
in the limit $\delta\to 0$ the ratio
$c_2/c_1$ equals $2$ (not $1$ as for conventional weak disorder)
inside the band (it vanishes outside the band). 
This energy-insensitivity has encouraged 
the authors of Ref.\ \onlinecite{DLA}
to conclude that single-parameter scaling is fulfilled, and
to introduce a novel criterion for single-parameter scaling.
As we will discuss now, these findings do not carry over to 
the fluctuations beyond the variance, characterized by $c_k$ with $k\geq 3$.

\begin{table}[t]
\begin{ruledtabular}
\begin{tabular}{c|c|c|c}
\quad 
& $D^{2/3}t^{-1/3}\ll\ep\ll t$ 
& $D/t\ll|E|\ll t$    
& $\ep,|E| \sim t$
\\ \hline
$c_1, c_2$
& $D/(4 t \ep)$ 
& $D/(4 t^2)$ 
& $D/(4t^2-E^2)$
\\ 
$c_3, c_4$
& $33 D^3/(128\, \ep^4 t^2)$ 
& $9D^3/(32\, E^2t^4)$ 
& $\propto D^3/t^6$
\\ 
$c_5, c_6$
& $5175 D^5/(2048\,\ep^7 t^3)$ 
& $135 D^5/(128\, E^4 t^6)$ 
& $\propto D^5/t^{10}$
\\ 
\end{tabular}
\end{ruledtabular}
\caption{The leading asymptotic values of the coefficients
$c_k$ in the case of weak Gaussian disorder $D\ll t^2$
upon the deviation $\ep=2t-|E|$ from the band edge 
(second column) or from the band center (third column). 
These results are obtained from the saddle-point
analysis of Eq.\ (\ref{basic}). \cite{ST2002,ST2003a}
The last column represents the 
generic values inside the band.}
\label{tab:Casymptotic}
\end{table}

The coefficient $c_3$ can be found from Eq.\ (\ref{d}) as
\beml
\label{c3}
\beq
c_3\!\!&=&\!\!3\!\int_{-\infty}^\infty\!\!\!\!\!dz
\ln\lt|\frac{s}{z-1/s}\rt|
\lt[c_2-\ln^2\lt|\frac{s}{z}\rt| 
\rt]h_0(z)\n\\
\!\!&+&\!\!3
\!\int_{-\infty}^\infty\!\!\!\!\!dz
\ln\lt|\frac{s}{z}\rt|
\lt[\ln\lt|\frac{s}{z}\rt|-2\ln\lt|\frac{s}{z-1/s}\rt| 
\rt]\Sigma(z),\qquad\\
\Sigma(z)\!\!&=&\!\!\frac{1}{2\pi}\im\!
\s_{n=1}^\infty
\lt[\frac{1}{z\!-\!s}-\frac{1}{z\!-\!p_n(s^\ast)}
\rt]\ln\frac{s^\ast}{p_n^{-1}(z)},
\eq
\eml
where the function $p_n^{-1}(z)$ stands for the inverse 
of $p_n(z)$. 
The ratio $c_3/c_1$ is plotted at Fig.\ (\ref{fig:ratio})
versus the energy. The plot
clearly displays a sequence of sharp dips,
which appear exactly at energies 
$E=-2 t\cos(\pi p/q)$ where $p$ and $q$ are integer, and become more narrow 
in the limit $\delta\to 0$. The anomaly in the band 
center is the biggest one and reaches 
about $3$\% of the absolute value of the ratio $c_3/c_1$
in the limit $\delta\to 0$. The existence of such anomalies
for the inverse localization length $\xi^{-1}=c_1$
has been pointed out by Lambert, \cite{Lambert}
but for this quantity they only show up
in higher orders of the expansion in the disorder strength, with
exception of the band edge $|E|=2$ 
and the band center $E=0$. \cite{Lambert,Derrida,KW}
For conventional weak disorder with $D\ll t^2$ 
the other anomalies should be seen in the higher 
coefficients $c_k$ with $k\geq 3$. However, those
cumulants are themselves suppressed by orders of $D/t^2$ 
(see Table \ref{tab:Casymptotic}), again with the exception of
the band edge and the band center, where they are of the
same order as $c_1$ and $c_2$. \cite{ST2002,ST2003a,ST2003b,DEL}

\begin{figure}[t]
\includegraphics[width=8cm]{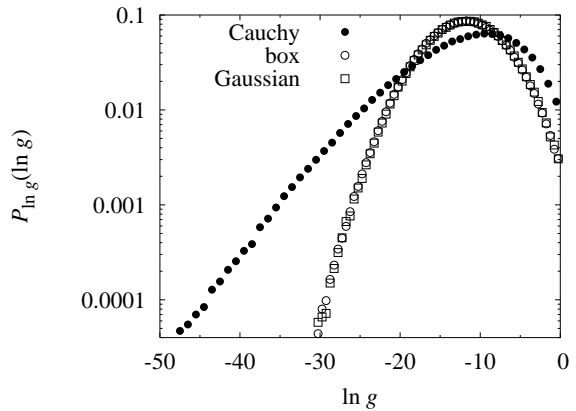}
\caption{Distribution function $P_{\ln g}(\ln g)$
obtained from the Anderson model
(\ref{AM}) with Cauchy disorder (full circles),
box disorder (open circles), and Gaussian disorder (open squares).
Parameters  are in the localized regime $L/\xi=6$,
with localization length $\xi=200$ lattice constants,
and $E=0.2 t$.
In this semilogarithmic plot, the lognormal distribution function
for conventional weak (box or Gaussian) disorder maps to an inverted parabola,
while  the straight-line asymptotics found for Cauchy disorder correspond to
a power-law tail in $P_g(g)$.
}
\label{fig:3}
\end{figure}

In striking contrast,
in the Lloyd model the coefficients $c_k$ increase very rapidly with
increasing index $k$.  
In the limit $\delta \to 0$
we indeed observe $c_2/c_1=2$, $c_3/c_1=5$, 
$c_4/c_1\approx 20$, $c_5/c_1\approx 100$
for $p/q$ irrational. 
The analysis of Eqs.\ (\ref{TheSet}) for the Lloyd model demonstrates that 
the generating function $\mu(\lambda)$ exists only for $\lambda <\lambda_c$, 
where the convergence radius $\lambda_c<1$, which
implies a factorial growth of the ratios $c_k/c_1$ for large $k$. 
Such a behavior is consistent with a power-law tail in
the conductance distribution function
$P_g(g)\sim  g^{-(2-\lambda_c)/2}$ for $g\to 0$.
For a general power law 
$P_V(V)\propto |V|^{-\beta}$ for $|V|\to\infty$,
$\lambda_c<\beta-1$ must be expected to depend on $\beta$, 
implying that the
precise form of the tail in $P_g(g)$ is not universal.
The power-law tail in $P_g(g)$ is certified by the numerical result for the
conductance distribution function for Cauchy disorder,
shown for $L/\xi=6$ in Fig.\ \ref{fig:3}.
The probability to find a vanishing conductance in the Lloyd model 
is strongly enhanced as compared to the case of 
conventional weak disorder, which displays log-normal tails.
In this case, the generating function $\mu(\lambda)$ 
is well-defined for all $\lambda$ in the whole energy range.
Moreover, far from the band edges ($\ep\equiv 2t-|E| \gg D^{2/3}t^{-1/3}$)
and far from the band center ($E\gg D/t$),
it acquires a universal parabolic form
$\mu(\lambda)=\xi^{-1} (\lambda+\lambda^2/2)$, since
$c_1=c_2=D/(4t^2-E^2)$, while all other coefficients can be disregarded 
in the limit $D\ll t^2$ (see Table \ref{tab:Casymptotic}). 

In conclusion, we have studied analytically 
the statistics of localized wave functions
in the Lloyd model, 
which is frequently used to analyze dynamical localization.
We have found that 
the distribution functions of the conductance $g$ and 
of the local density of states $\nu$ do not have a
log-normal form.
Moreover, even in the limit of vanishing disorder
these  distribution functions reveal sharp anomalies 
at energies $E=-2 t\cos(\pi p/q)$, 
with $p/q$ a rational number. These specific features
can be attributed to the power-law decay of the disorder 
distribution function.
They sensitively affect the moments (including
fractional moments) of $g$ and $\nu$ and
demonstrate that for such distribution functions
the wave-function statistics are highly non-universal.
It would be striking to see similar effects
for the prelocalized states in the diffusive regime
$L\lesssim \xi$ of multichannel systems. \cite{AKL}

We thank B.~L.~Altshuler and S.~Fishman for 
illuminating discussions.


\begin{thebibliography}{99}
\bibitem{footnote}
The kernel function $K(z,z')$ 
is not uniquely defined, since one can always add to it
an arbitrary function depending only on $z$. The fact that
such transformations do not affect the results
ensures the independence of the coefficients $c_k$ on the conditions 
of the initial-value problem.
\end{thebibliography}
\end{document}